\documentclass{emulateapj}

\usepackage{apjfonts}
\usepackage{graphicx}

\slugcomment{ApJ in press}

\begin{document}
\title{On the Eccentricity Distribution of Exoplanets from Radial Velocity Surveys}

\shorttitle{ECCENTRICITY DISTRIBUTION OF EXOPLANETS}

\shortauthors{SHEN \& TURNER}

\author{Yue Shen\altaffilmark{1} and Edwin L. Turner\altaffilmark{1,2}}
\affil{$^1$Princeton University Observatory, Princeton, NJ 08544,
USA\\
$^2$Institute for the Physics and Mathematics of the Universe,
University of Tokyo, Kashiwa 277-8568, Japan}
\email{yshen@astro.princeton.edu; elt@astro.princeton.edu}

\begin{abstract}
We investigate the estimation of orbital parameters by
least-$\chi^2$ Keplerian fits to radial velocity (RV) data using
synthetic data sets. We find that while the fitted period is
fairly accurate, the best-fit eccentricity and $M_p\sin i$ are
systematically biased upward from the true values for low
signal-to-noise ratio $K/\sigma\lesssim 3$ and moderate number of
observations $N_{\rm obs}\lesssim 60$, leading to a suppression of
the number of nearly circular orbits. Assuming intrinsic
distributions of orbital parameters, we generate a large number of
mock RV data sets and study the selection effect on the
eccentricity distribution. We find the overall detection
efficiency only mildly decreases with eccentricity. This is
because although high eccentricity orbits are more difficult to
sample, they also have larger RV amplitudes for fixed planet mass
and orbital semi-major axis. Thus the primary source of
uncertainties in the eccentricity distribution comes from biases
in Keplerian fits to detections with low-amplitude and/or small
$N_{\rm obs}$, rather than from selection effects. Our results
suggest that the abundance of low-eccentricity exoplanets may be
underestimated in the current sample and we urge caution in
interpreting the eccentricity distributions of low-amplitude
detections in future RV samples.
\end{abstract}
\keywords{planetary systems -- techniques: radial velocities}

\section{Introduction}
With the rapid increase in the rate of exoplanet detections, it
has become feasible to study their statistical properties (for a
recent review, see Udry \& Santos 2007). The distributions of
their orbital parameters and correlations with host star
properties are crucial for our understanding of planet formation.
Up to Feb 2008, over 200 exoplanets have been announced, most of
which were detected by the radial velocity (RV) technique. Among
the parameters that can be derived from RV data, the orbital
eccentricity has a somewhat unexpected distribution, with an
extended tail of high eccentricities ($e\gtrsim 0.1$). Although
the exact form of this eccentricity distribution is still
uncertain to some extent, especially at the lowest eccentricities
(i.e., comparing the distribution in Butler et al. 2006 and in the
most recent catalog), it is clear that this distribution is quite
different from that of the Solar system. There have already been
several theoretical attempts to explain such an eccentricity
distribution (e.g., Tremaine \& Zakamska 2004 and references
therein; Juric \& Tremaine 2007; Ford et al. 2007; Zhou et al.
2007).

However, the largest current exoplanet catalog (e.g., Bulter et
al. 2006 and their updates, hereafter Bulter06) is by no means
homogeneous. Survey strategies, selection biases in the RV
technique and uncertainties in best-fit orbital solutions could
all bias the intrinsic distributions of orbital parameters,
especially when taken together. Eccentric orbits have larger
amplitudes than circular orbits when holding other parameters
fixed, while failure to (time) resolve the perihelion approach can
lead to non-detection for an eccentric orbit system. These two
effects are believed to roughly cancel but detailed simulation is
needed (e.g., Endl et al. 2002; Cumming 2004). It is also well
known that the errors in least-$\chi^2$ Keplerian solutions become
asymmetric for noisy data (e.g., Ford 2005; Butler et al. 2006),
which is especially true for eccentricity. On the other hand, for
circular orbits the eccentricity can only be scattered upward by
error, so an obvious bias exists for circular orbits. As the RV
surveys progress, more and more low amplitude (low signal-to-noise
ratio) detections will be reported. Significant uncertainties
remain in the distributions of orbital parameters for those
exoplanets using best-fit orbital solutions. There are already
some cases where the orbit eccentricity is poorly constrained by
the RV data (e.g., Jones et al. 2006).

The purpose of the paper is to explore processes that might
distort measurements of the shape of the intrinsic eccentricity
distribution using synthetic data. First we investigate the
reliability of single Keplerian fits to mock RV data sets as
function of signal-to-noise ratio, and see if any bias arises from
Keplerian fitting to noisy data (\S\ref{sec:kep_fit_eff}). Second,
we construct statistical models of orbital parameter distributions
and pass them through a simulated detection pipeline, in order to
see if there is any serious selection effect on eccentricity
(\S\ref{sec:stat}). We discuss our results and apply them to the
current RV planet sample in \S\ref{sec:diss}, and we summarize our
findings in \S\ref{sec:conc}.

\begin{figure*}
  \centering
    \includegraphics[width=0.95\textwidth]{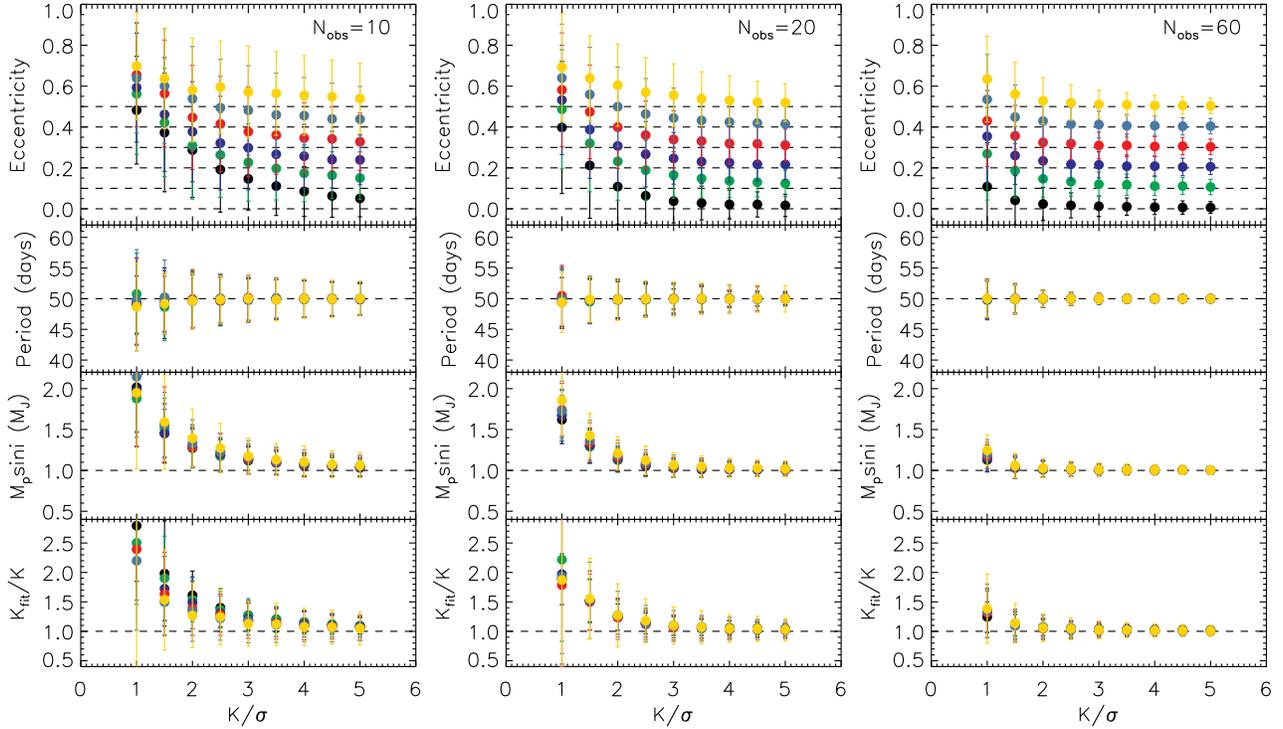}
    \caption{Effects of signal-to-noise ratio and number of observations on the best-fit orbital
    parameters. Filled circles are median values; the true orbital parameters are
    denoted as horizontal lines (see the text for details), where $\omega$
    and temporal offset $t_0$ are chosen at random. Error bars show standard
    deviations. From left to right, $N_{\rm obs}=10,20,60$.}
    \label{fig:SN_eff}
\end{figure*}

\section{Keplerian Fitting to Mock RV Data}
\label{sec:kep_fit_eff}

We consider single planet detections throughout the paper.
Assuming a true orbit (period $P$, eccentricity $e$, minimal
planet mass $M_p\sin i$, argument of periastron $\omega$, and
temporal offset $t_0$) and the host star mass $M_*$, we generate
mock RV data sets with Gaussian errors $\sigma$ as perturbations.
For simplicity $\sigma$ is assumed to be the same for all data
points in a set, but we relax this assumption later
(\S\ref{sec:diss}). In practice, $\sigma$ comes from RV
measurement errors, stellar jitter and possible existence of
multiple planets (e.g., Ford 2006)\footnote{We note that in
general the additional reflex motion induced by unknown secondary
planets cannot be treated as independent perturbations. Second,
for simplicity we do not consider secular trends in the orbital
motion of the primary planet induced by unseen companions, which
are typically on much longer timescales than the period and
observation time baseline considered here.}. Following Cumming
(2004), we define the signal-to-noise ratio ${\cal R}=K/\sigma$
where $K$ is the semi-amplitude of the theoretical RV curve, and
in the limit of $M_p\ll M_*$ we have (e.g., Cumming et al. 1999),
\begin{equation}\label{eqn:amp}
K=\bigg(\frac{2\pi G}{P}\bigg)^{1/3}\frac{M_p\sin
i}{M_*^{2/3}}\frac{1}{\sqrt{1-e^2}}\ .
\end{equation}
Here the amplitude $K$ is defined in terms of the intrinsic
properties of a planet's orbit, since we ``know'' the true orbital
parameters in advance.

\begin{figure*}
  \centering
    \includegraphics[width=0.45\textwidth]{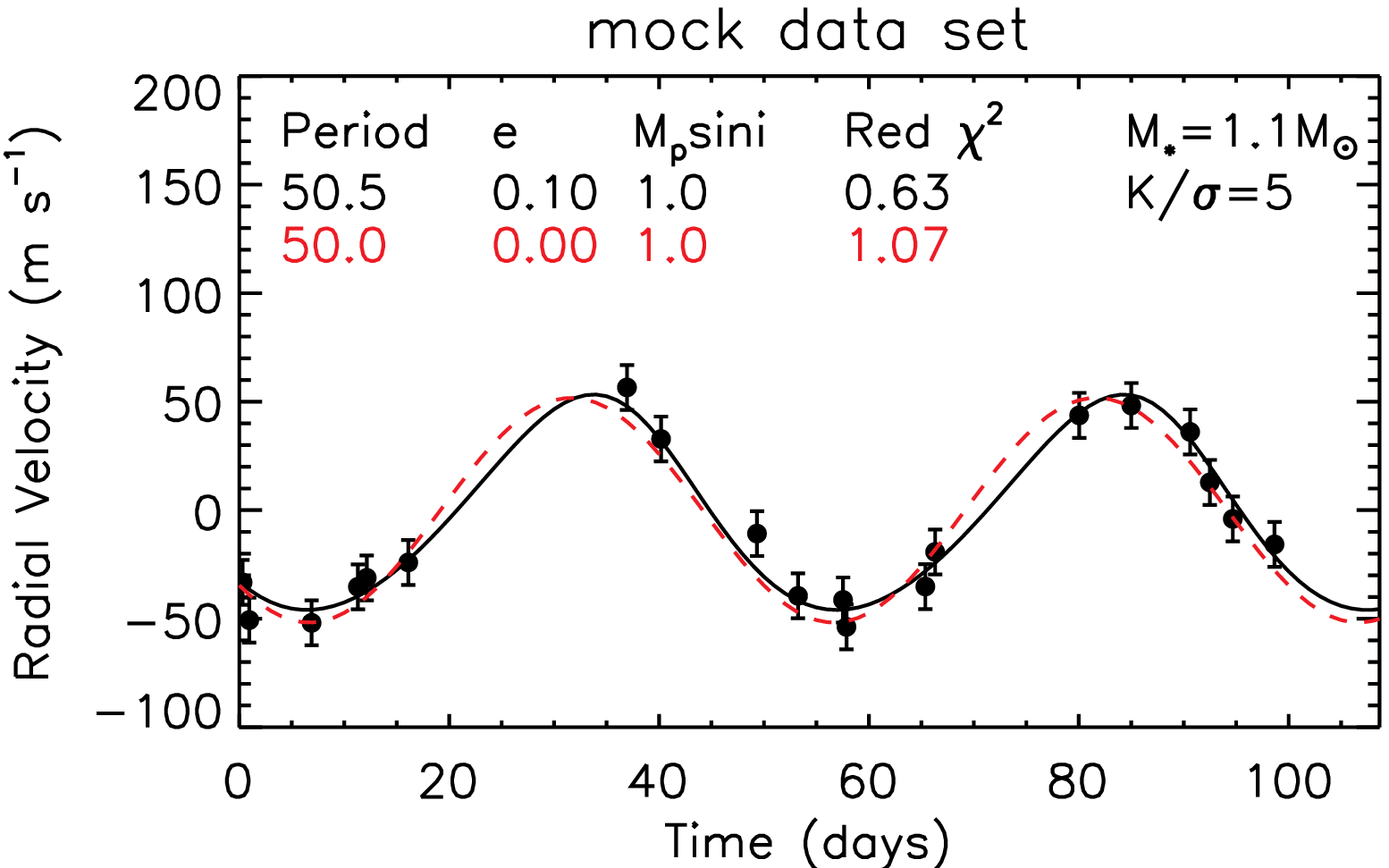}
    \includegraphics[width=0.45\textwidth]{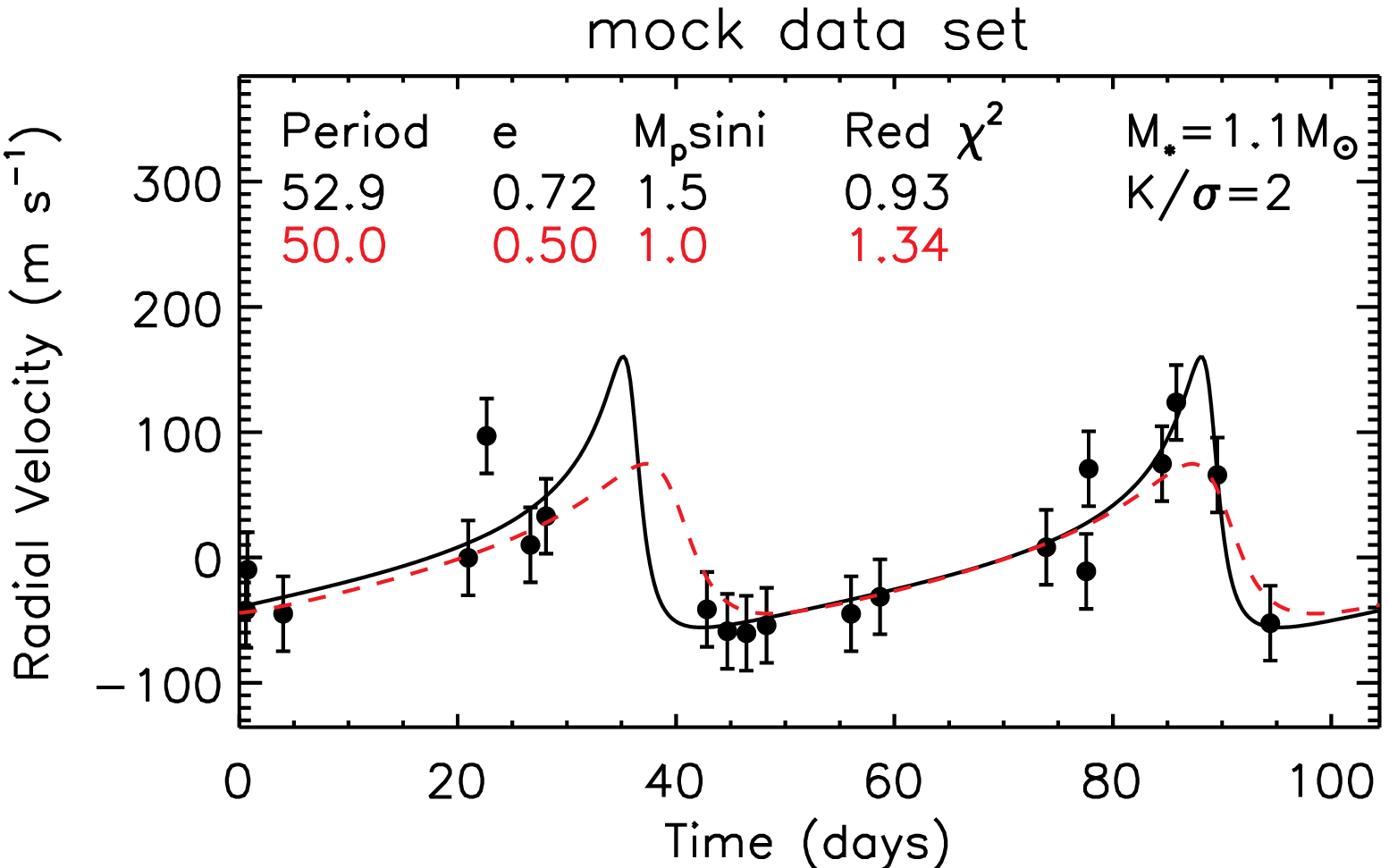}
    \caption{{\em Left}: An example of Keplerian fit to mock data with ${\rm
    K/\sigma}=5$ and $N_{\rm obs}=20$
    where $e_0=0$. The perturbed RV data prefer an eccentric orbit over the original
    circular orbit; but the fitted period and $M_p\sin i$ are fairly accurate. {\em Right}:
    An example of large eccentricity orbit fits with $K/\sigma=2$ and $N_{\rm obs}=20$ where $e_0=0.5$. In both cases the dashed lines
    are the RV curves expected from the true orbital parameters while the solid lines are best-fit Keplerian orbits to the data.}
    \label{fig:example1}
\end{figure*}

In addition to the signal-to-noise ratio ${\cal R}$, another
important quantity is the number of observations $N_{\rm obs}$.
Clearly the larger ${\cal R}$ and $N_{\rm obs}$ are, the better
the quality of the data set.

We consider planets where the orbital period is shorter than the
observational time span. Orbits with periods longer than the time
baseline are difficult to detect due to limited sampling and lower
amplitude (e.g., Cumming 2004). Even if detected, the best-fit
orbital solution would often be very different from the true one
(Shen et al. 2008, in preparation). The observation times are
randomly generated within the observational time span, which
somewhat mimics the effects of realistic radial velocity sampling.

In Fig. \ref{fig:SN_eff} we plot the relations of fitted
eccentricity, period, minimal planet mass and semi-amplitude to
their true values, as functions of ${\cal R}$, for a specific case
with $M_*=1.1\ M_\odot$, $P=50$ days, $M_p\sin i=1\ M_J$, and
eccentricity values $e=0.0-0.5$ at intervals of $\Delta e=0.1$.
Each set of true orbits was used to generate 5000 mock RV data
sets with randomly chosen $\omega$ and temporal offset $t_0$, and
with Gaussian noise determined from the assumed signal-to-noise
ratio. We used the Levenberg-Marquardt method (e.g., Press et al.
1992) to minimize $\chi^2$, and the true orbital parameters were
used as an initial guess for the $\chi^2$ fitting. Using the true
orbital parameters as the initial guess greatly speeds up the
convergence and increases the rate of successful fits. In the
realistic case when we do not know the actual orbital parameters,
the success of a fit also depends on an appropriate initial guess.
We will come back to this point in \S\ref{sec:stat}. The mock
observational time span $T$ was set to be two true periods, but we
find almost identical results as long as the period is shorter
than the time span. Finally, we simulated three values of $N_{\rm
obs}=10,20,60$.

Not all attempts to find a Keplerian fit will result in a
detection. In an actual radial velocity search for planets, some
threshold on the false alarm probability (FAP), the probability
that a signal power will arise purely from noise, of a Keplerian
fit, is normally used. Throughout the paper we adopt a fiducial
${\rm FAP}=10^{-2}$, which is essentially always satisfied for
large $K/\sigma$ and $N_{\rm obs}$. There are several ways to
estimate the FAP using Monte Carlo simulations or analytical
formulae. For simplicity we have followed the analytical procedure
described in Cumming (2004) to estimate the FAP: Given the
least-$\chi^2$ value $\chi^2_{\rm Kep}$ from the Keplerian fit,
and the least-$\chi^2$ value $\chi^2_{\rm mean}$ from the fit of a
constant to the data, we calculate a power $z_0$ using eqn. (7) in
Cumming (2004). An estimate of the FAP is then eqn. (5) in Cumming
(2004): ${\rm FAP}=1-[1- {\rm Prob}(z>z_0)]^{\cal M}$, where
${\cal M}\approx T/T_{\rm min}$ is the number of independent
frequencies with $T$ the observational time span and $T_{\rm
min}=2$ days the lower bound of period searched in the Keplerian
fitting, and ${\rm Prob}(z>z_0)$ is the probability distribution
given by eqn. (8) in that paper.

All three panels in Fig. \ref{fig:SN_eff} show that the fitted
period is fairly accurate. However, for eccentricity and $M_p\sin
i$ (and also the amplitude $K$ following eqn. \ref{eqn:amp}) the
{\em median} fitted values are biased upward, and this behavior
remains for true eccentricities as large as $e_0\sim 0.8$ in our
simulations. This is not a failure of the $\chi^2$ minimization
method. In Fig. \ref{fig:example1} we show two examples of
Keplerian fits where both RV data sets prefer a more eccentric
orbit. The biases in $e$ and $M_p\sin i$, as well as the scatter
around the median, decrease when ${\cal R}$ or $N_{\rm obs}$
increases, as would be naively expected.

These fitting biases result from the fact that, in general, the
least-$\chi^2$ solution is {\em not} an unbiased estimator of the
true parameters for the non-linear Keplerian model even if the
data errors are Gaussian.  In particular, the probability
distribution ${\cal P}(\mathbf{\hat{\theta}}|\mathbf{\theta})$ is
not Gaussian, where $\mathbf{\theta}$ is the set of true parameter
and $\mathbf{\hat{\theta}}$ is the best-fit orbital solution; and
it only approaches a Gaussian distribution when the errors are
small.

To fully take account these biases we need to know ${\cal
P}(\mathbf{\hat{\theta}}|\mathbf{\theta})$ for different
combinations of $\mathbf{\theta}$, $K/\sigma$ and $N_{\rm obs}$,
which can be explored numerically as in Fig. \ref{fig:SN_eff}.
However, for low intrinsic eccentricities ($e_0\lesssim 0.5$), the
apparent fitting bias in eccentricity can be quantitatively
understood as follows. While the error in the fitted eccentricity
$\sigma_e$ is not Gaussian when $\sigma_e$ is comparable to $e$,
the fitted $x\equiv e\cos\omega$ and $y\equiv e\sin\omega$
preserve Gaussianity much better provided that the intrinsic
eccentricity is small (e.g., Butler et al. 2006). Assuming their
fitted values are Gaussian distributions around true values
$x_0\equiv e_0\cos\omega_0$ and $y_0\equiv e_0\sin\omega_0$ with
dispersion $\sigma_{0}$, the probability distribution of the
fitted eccentricity $e$ is given by
\begin{eqnarray}\label{eqn:ana_bias}
{\cal P}(e)&\propto& \int
\exp\bigg[-\frac{(x-x_0)^2}{2\sigma_0^2}-\frac{(y-y_0)^2}{2\sigma_0^2}\bigg]\delta(e-\sqrt{x^2+y^2})dxdy\nonumber  \\
&\propto&
e\exp\bigg(-\frac{e^2}{2\sigma_0^2}\bigg)I_0\bigg(\frac{ee_0}{\sigma_0^2}\bigg)\
,
\end{eqnarray}
where $I_0$ is the modified Bessel function of the first kind. For
circular orbits ($e_0=0$) it is a Rayleigh distribution; for
eccentric orbits and at the low-dispersion limit ($\sigma_0\to 0$)
it is approximately a Gaussian distribution.

\begin{figure}
  \centering
    \includegraphics[width=0.45\textwidth]{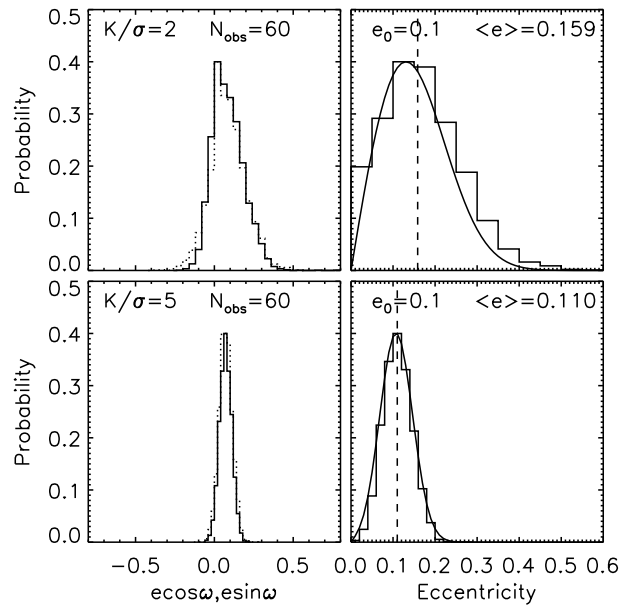}
    \caption{Examples of fitted distributions becoming non-Gaussian when
    the signal-to-noise ratio decreases (see text for the intrinsic parameters).
    {\em Left}: Distributions of fitted $e\cos\omega$ (solid histogram) and
    $e\sin\omega$ (dotted histogram). {\em Right}: Distributions of fitted eccentricities.
    The median value of fitted
    eccentricities, denoted as dashed vertical lines, is biased from the intrinsic
    value $e_0=0.1$ in the $K/\sigma=2$ case. The difference between
    predictions (solid lines)
    from eqn. (\ref{eqn:ana_bias}) and the simulated distributions is due to the fact that
    the distributions of $e\cos\omega$ and $e\sin\omega$ are not precisely Gaussian.}
    \label{fig:SN_test}
\end{figure}

Fig. \ref{fig:SN_test} shows an example of $[P,e_0,M_p\sin i,
\omega_0,t_0,M_*]=[50, 0.1, 1, \pi/4,10,1.1]$ with $N_{\rm
obs}=60$. For a high signal-to-noise ratio, $K/\sigma=5$ (lower
panels), both distributions of fitted eccentricity and
$e\cos\omega$ ($e\sin\omega$) are approximately Gaussian. For a
low signal-to-noise ratio, $K/\sigma=2$ (upper panels), the
eccentricity distribution significantly deviates from a Gaussian
while $e\cos\omega$ and $e\sin\omega$ are still distributed in an
approximately Gaussian form. We plot the predictions of fitted
eccentricity distribution from eqn. (\ref{eqn:ana_bias}) as solid
lines, using the best-fit Gaussian dispersion of the
$e\cos\omega$/$e\sin\omega$ distributions. In this way we can also
understand the upward median biases even at intrinsic
eccentricities as high as $\sim 0.8$, although the fitted
$e\cos\omega$ and $e\sin\omega$ are no longer Gaussian at such
high eccentricities--our simulations themselves are more revealing
in this case.

A conservative empirical criterion (based on our numerical experiments)
for a {\em median} bias in
fitted eccentricity $\left(\langle e\rangle-e_0\right)\lesssim
0.05$ is given by the following joint constraint on $K/\sigma$ and
$N_{\rm obs}$:
\begin{equation}\label{eqn:bias}
(K/\sigma)N_{\rm obs}^{1/2}\gtrsim 15\ , %\ge 13.3(1.25-e)^{0.588}\.
\end{equation}
which works reasonably well for $10\le N_{\rm obs}\le 200$. This
is also an approximate criterion that the {\em median} bias in
$M_p\sin i$ is less than $10\%$. Note again here that $K$ is
defined using intrinsic planet properties. In practice, $K$ can
only be estimated from the best-fit solution and will on average
be overestimated a bit at the low-amplitude end, according to the
fitting biases discussed here (e.g., Fig. \ref{fig:SN_eff}). If
using $K_{\rm fit}$ instead, we found a good approximation is to
replace equation (\ref{eqn:bias}) with
\begin{equation}\label{eqn:bias2}
(K_{\rm fit}/\sigma)N_{\rm obs}^{1/2}\gtrsim 17\ ,
\end{equation}
which compensates the fitting bias of amplitude at the low
$K/\sigma$ end.

\section{Statistical Models of Orbital Parameter Distributions}
\label{sec:stat}

We now investigate the selection bias on eccentricity using
synthetic RV data sets generated from model distributions of
orbital parameters. The detailed comparison of different model
distributions and constraints from observational data will be
presented in a paper in preparation (Shen et al. 2008). Our
purpose here is to investigate the possible selection bias on
eccentricity, therefore essentially any well behaved model
distributions will do. Nevertheless, we choose a model with
physically plausible distributions, as we describe below.

\begin{figure}
  \centering
  \includegraphics[width=0.45\textwidth]{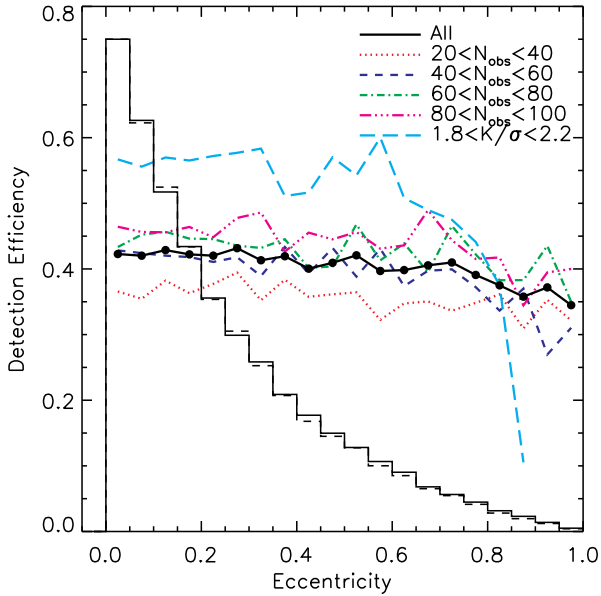}
  \caption{Colored lines show the detection efficiencies as a function of eccentricity within a given
    bin of $N_{\rm obs}$. The long-dashed cyan line show the detection efficiency for fixed signal-to-noise ratio. The solid
    and dashed histograms show the true eccentricity distributions for all the mock orbits and
    the detected orbits, with the same peak normalization.}
    \label{fig:DE}
\end{figure}

The period distribution is a log-normal with log mean of 10 yrs
and dispersion of 1 dex. The planet mass $M_p$ distribution is a
power-law $ndM_p\propto M_p^{-1}$ within $[0.1,100]\ M_J$; this
extends well into the ``brown dwarf desert'' due to the upper
limit of $100\ M_J$.  However such objects are too rare to have
any statistical impact on our results. The orientation of the
planet orbital plane is random, as is the argument of periastron
$\omega$ within $[0,2\pi]$ and the temporal offset $t_0$ within
one period. For the eccentricity distribution we chose a model
that peaks at $e=0$ and diminishes to zero at $e=1$:
\begin{equation}
{\cal P}(e)de\propto
\bigg[\frac{1}{(1+e)^a}-\frac{e}{2^a}\bigg]de\ ,
\end{equation}
and we assume $a=4$ for this specific study. Our choice of this
particular model distribution is because the peak of the {\em
observed} eccentricity distribution has shifted from $e\approx
0.2$ to $e\approx 0$ using the latest exoplanet catalog\footnote{
http://www.exoplanets.org}. This model distribution of
eccentricity is shown as the solid histogram in Fig. \ref{fig:DE}.
Finally we specify an observational time baseline of $T=10$ years
and an overall radial velocity error $\sigma=10\ {\rm m\, s^{-1}}$
for each simulated observation; the number of observations of each
object is randomly distributed from $N_{\rm obs}=20$ to $N_{\rm
obs}=100$. These parameters are intended to be typical of
currently RV surveys (e.g., Cumming et al. 2008).

We generate 100,000 mock orbits of the model distributions and
pass them through the Keplerian detection pipeline with a
detection threshold ${\rm FAP}=0.01$. For the initial guess of
orbital solution, we again take advantage of our mock data sets by
using the true orbital parameters. In the realistic case, the
initial guess for period can be identified using traditional
Lomb-Scargle periodogram (Lomb 1976; Scargle 1982), but
appropriate initial guesses for $[e,\omega,t_0]$ are crucial for
the convergence of Keplerian fits for orbits with high
eccentricities, where the fit can easily be trapped in a local
minimum due to the complicated $\chi^2$ space or fail to converge
at all, and hence the planet is undetected. We have run a
substantial number of tests, where we tried a fine grid of
$[e,\omega,t_0]$ as initial guesses in each Keplerian fit. While
time consuming, we found that in the vast majority cases the
Keplerian fits find the same minimum as using the true parameters
as the initial guess. Serendipitous aliasing cases do exist, where
the global minimum solution deviates catastrophically from the
true orbital parameters, and which are for the most noisy data
sets. Hence in the realistic case, special care must be taken to
find the global minimum efficiently.

In Fig. \ref{fig:DE} we show the overall detection efficiency (the
fraction of detected orbits) as function of eccentricity in filled
circles, and different line types represent the detection
efficiency when binned in $N_{\rm obs}$. Clearly increasing
$N_{\rm obs}$ increases the detection efficiency. However, the
overall detection efficiency only slowly decreases with
eccentricity. This is because although high eccentricity orbits
are more difficult to sample, they also have larger amplitude $K$
on average. When we plot the detection efficiency with constant
amplitude $K/\sigma\approx 2$, we see a decrease for $e\gtrsim
0.6$, as shown by the cyan long-dashed line in Fig. \ref{fig:DE},
which is consistent with previous studies (e.g., Endl et al. 2002;
Cumming 2004). Therefore, there is no strong selection bias
against eccentricity, as demonstrated by the similar (true)
eccentricity distributions of the mock orbits and the detected
ones in Fig. \ref{fig:DE}. Note, however, the absolute value of
the overall detection efficiency here depends on the model
distributions and the quality of the survey (noise level and
number of observations).

\begin{figure}
  \centering
    \includegraphics[width=0.45\textwidth]{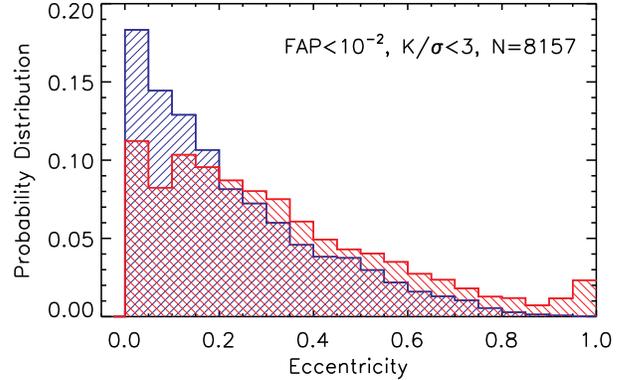}
    \caption{Low $K/\sigma$ detections for the statistical model (8157 planets detected). Higher and lower histograms show the distributions
    of the intrinsic and best-fit eccentricities of these detected planets. Shown here is the bias
    caused by Keplerian fitting: $\sim 40\%$ of the nearly-circular ($e\le 0.1$) orbits are erroneously assigned to
    more eccentric orbits by their best fits.}
    \label{fig:stat_model1_low_SN}
\end{figure}

\begin{figure*}
  \centering
    \includegraphics[width=0.8\textwidth]{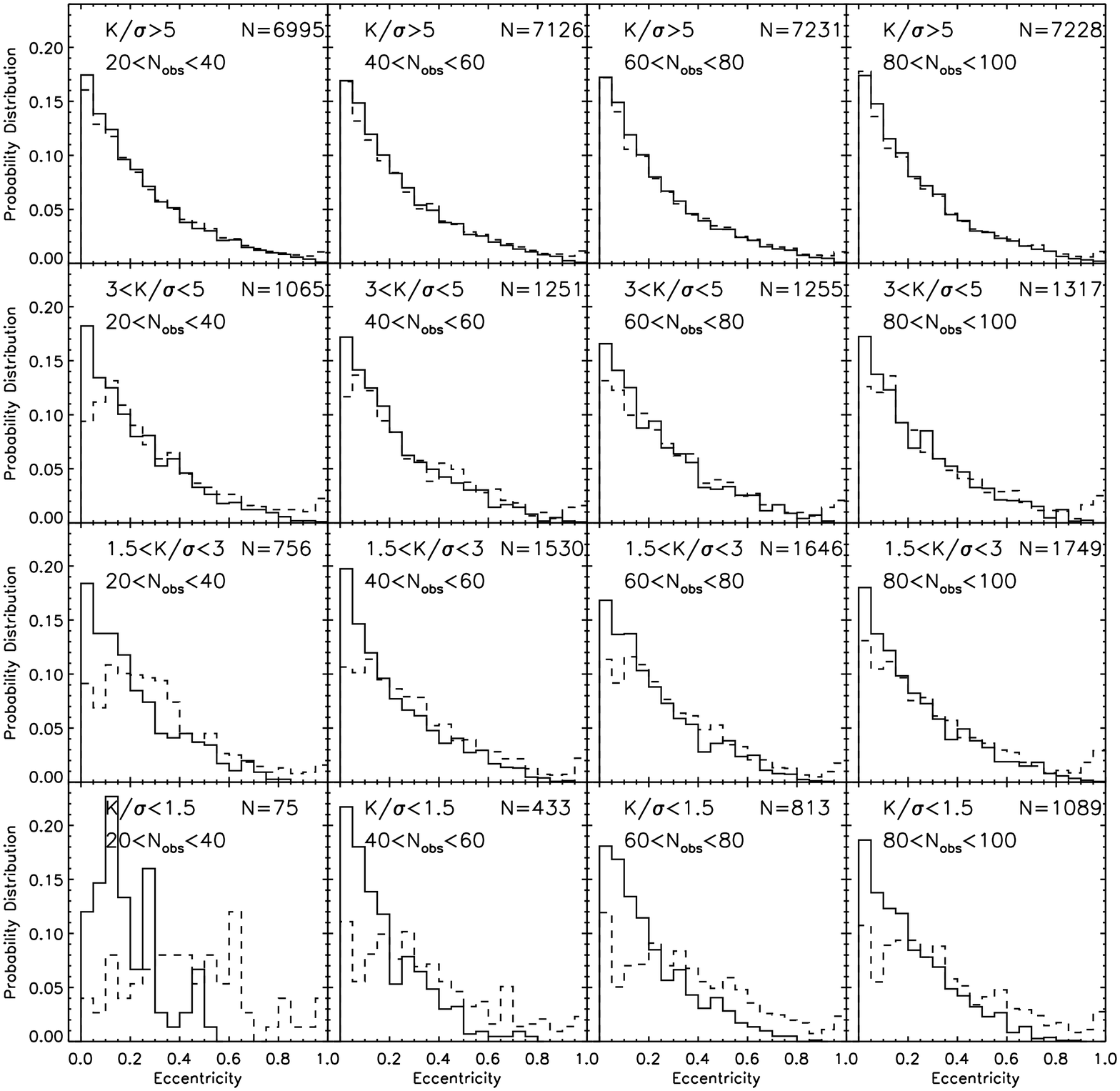}
    \caption{The effects of Keplerian fitting on the eccentricity distributions. The solid and
    dashed histograms are the distributions of intrinsic and fitted eccentricities in different
    $K/\sigma$ and $N_{\rm obs}$ bins. All planets are detected by the Keplerian fitting pipeline, and the
    numbers of detected planets are listed at the top-right corner of each
    panel. Those bins with the lowest $K/\sigma$ and $N_{\rm obs}$
    are the most seriously affected by the fitting bias.
    }
    \label{fig:stat_e_bias}
\end{figure*}

However, as we have shown in \S\ref{sec:kep_fit_eff}, $\chi^2$
Keplerian fits will have non-negligible effects for low amplitude
detections and therefore bias the underlying eccentricity
distribution. Although the detection efficiency decreases rapidly
when $K/\sigma$ decreases, there are also many more objects in the
low amplitude regime given our model distributions and noise
level. We show this bias effect in Fig.
\ref{fig:stat_model1_low_SN}, where the blue and red histograms
show the distributions of true and fitted eccentricities for low
amplitude detections with $K/\sigma<3$. The small pile up of
fitted $e\gtrsim 0.9$ orbits is caused by detections with period
longer than the observation time where the Keplerian fits failed
badly (Shen et al. 2008, in preparation). Other than that, the
overall effect of the Keplerian fitting bias is the transfer of
true low-eccentricity orbits to fitted high-eccentricity ones,
therefore distorting the underlying eccentricity distribution. As
expected from our simulations in \S\ref{sec:kep_fit_eff}, this
effect diminishes when we shift to high amplitude and/or large
$N_{\rm obs}$ detections, as shown in Fig. \ref{fig:stat_e_bias}.

\begin{figure}
  \centering
    \includegraphics[width=0.45\textwidth]{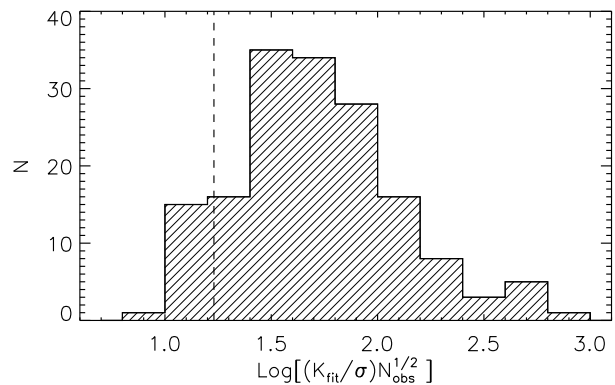}
    \caption{Inspection of the exoplanet catalog as of January 26, 2008 with $P<20$ days planets excluded.
    The histogram shows the distribution of $(K_{\rm fit}/\sigma)N_{\rm
    obs}^{1/2}$ for these planets. We have used the RMS residual of each model fit as an estimate for $\sigma$. The vertical dashed line is at $(K_{\rm fit}/\sigma)N_{\rm
    obs}^{1/2}=17$, which is the approximate bound below which the best-fit eccentricity is
    affected by the fitting bias (e.g., eqn. \ref{eqn:bias2}).}
    \label{fig:butler_cat_bias}
\end{figure}

\begin{figure}
  \centering
    \includegraphics[width=0.45\textwidth]{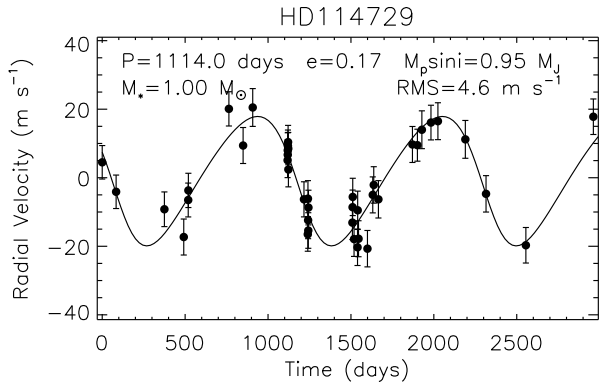}
    \caption{An example of the Kepler fits to RV data in the Butler06 catalog (HD 114729b). This particular case has
    $N_{\rm obs}=42$ and median $K_{\rm fit}/\sigma\approx 3.5$, where $\sigma$ is estimated from the
    published RV measurement errors and stellar jitter.}
    \label{fig:butler_cat}
\end{figure}

\section{Discussion}
\label{sec:diss}

Our results lead naturally to the question of how much bias in the
eccentricity distribution is to be expected in the Butler et al.
(2006) catalog. To test this we take their updated catalog as of
January 26, 2008. We exclude those planets with orbital periods
less than 20 days (i.e., those probably have been tidally
circularized). Over half of the planets in that catalog do not
have published RV data and measurement errors/stellar jitter
information, so we use the RMS scatter of the model fit as an
estimate of the noise $\sigma$. We also take the amplitude $K$ as
the best-fit value $K_{\rm fit}$. There are $162$ planets with
$K_{\rm fit}$, ${\rm RMS\ scatter}$ and $N_{\rm obs}$ available.
In Fig. \ref{fig:butler_cat_bias} we show the distribution of
$(K_{\rm fit}/\sigma)N_{\rm obs}^{1/2}$ defined in equation
(\ref{eqn:bias2}) for these planets. There are 19 planets with
$(K_{\rm fit}/\sigma)N_{\rm obs}^{1/2}<17$. Thus statistically
speaking, only $\sim 10\%$ of the planets are affected in the
current sample, which are predominately low-amplitude detections.

To test the effects of unequal errors and realistic sampling we
take HD 114729b in the Butler06 catalog as an example. It has a
best-fit eccentricity $0.167\pm0.055$ with $K_{\rm
fit}/\sigma\approx 3.5$ and $N_{\rm obs}=42$ where we have
estimated $\sigma$ using their published RV measurement errors and
stellar jitter. The actual RV data and their best-fit solution is
shown in Fig. \ref{fig:butler_cat}. Using the true time series and
data uncertainties derived from measurement errors and stellar
jitter for this system, but with simulated radial velocities, we
find that the probability of a fitted eccentric orbit with $e\ge
0.15$ arising from a circular orbit is $\sim 8\%$, and increases
to $20\%$ for a fitted $e\ge 0.1$, consistent with our results in
\S\ref{sec:kep_fit_eff} with constant uncertainties and random
sampling.

\section{Conclusions}
\label{sec:conc}

We have performed simulations of planet detection and orbital
parameter fits using mock radial velocity data sets. Two effects
that may affect the intrinsic eccentricity distribution are
considered: selection bias on eccentricity, and fitting bias in
the best-fit orbital solution.

We find that selection bias on eccentricity is negligible, as long
as the fitting routine is efficient in finding the global
solution. In a realistic survey, this requires a thorough search
in the parameter space, and/or with advanced algorithms optimized
for the search. Our finding is not in conflict with previous
studies (e.g., Endl et al. 2002; Cumming 2004), which claim that
the detection efficiency decreases for $e\gtrsim 0.6$. This is
because in previous studies, the detection efficiency as a
function of eccentricity is estimated at fixed $K$ while in our
study $K$ is larger for more eccentric orbits when other
parameters are fixed. On the other hand, we find that for
detections with low signal-to-noise ratio and small number of
observations, the best-fit eccentricity is biased upward in the
median value, which then gives rise to a change in the
eccentricity distribution from the intrinsic one.

Inspection of the current exoplanet sample shows only $\sim 10\%$
are likely to be significantly affected by the Keplerian fitting
bias. However, future radial velocity surveys will contain an
increasing number of low amplitude detections if the number of low
mass, large semi-major axis exoplanets grows as rapidly as
suggested by extrapolation of current results suggest (e.g., Udry
\& Santos 2007). When the sample of low amplitude detections (on
average less massive planets and more circular orbits) is large
enough for statistical study, the bias in the best-fit
eccentricity described here must be taken into account. Also, in
individual cases where an accurate estimate of eccentricity is
required, i.e., in modeling the habitable zone or tidal heating
issues, either high quality RV data or constraints from other
observations are required to support reliable conclusions. In the
mean time, there is also a need for a more statistically
sophisticated understanding of the uncertainties of derived
orbital solutions (e.g., Ford 2005; 2006).

Our results suggest that the intrinsic eccentricity distribution
may be even more peaked at $e\approx 0$ than the current observed
distribution. Some planet-planet scattering models tend to produce
a Rayleigh-distribution of eccentricities (e.g., Juric \& Tremaine
2007; Ford \& Rasio 2007) with reduced circular orbits, therefore
certain eccentricity damping mechanism such as interactions with a
protoplanetary disk may be required to reconcile these models with
observations.

\acknowledgements

We thank the anonymous referee for useful comments, as well as
Scott Tremaine and Mario Juri\'{c} for helpful discussions. This
research was supported in part by NASA grant NNG06GE27G.

\end{document}